\documentclass[prl,twocolumn,floatfix,showpacs]{revtex4}
\usepackage{graphics}
\usepackage{dcolumn}
\usepackage{bm}
\usepackage{epsfig}
\usepackage{pictex}

\begin{document}
\draft
\preprint{HEP/123-qed}
\title{Strong Coupling Theory 
of the Spinless Charges on the Triangular Lattices: 
Possibility of a New Quantum Liquid}
\author{Chisa Hotta
and Nobuo Furukawa}
\address{Aoyama-Gakuin University, 5-10-1, Fuchinobe, Sagamihara, Kanagawa 229-8558}
\date{\today}
\begin{abstract}
We propose a new type of charge liquid state 
in the spinless fermion system on a triangular lattice 
under strong inter-site Coulomb interactions, $V$. 
In the strong coupling limit ($t=0$), 
the ground state is classical and disordered 
due to geometrical frustration. 
The introduction of small $t$ will drive 
the system to a partially ordered phase which we call a "pinball liquid". 
A possibly long range ordered Wigner crystal solid coexist with 
a liquid component which are moving around them like a pinball. 
This liquid is dominant over wide range of filling, 
even away from the regular triangle, 
and is also realized in the hard core boson systems. 
Relevance to the organic $\theta$-ET$_2X$ is discsussed. 
\end{abstract}
\pacs{75.40, 75.50, 72.80.Sg, 72.80.Le}
\maketitle
\narrowtext
Geometrical frustration induces exotic states 
in the strongly correlated systems 
due to the severe competition among ordered states. 
They are explored experimentally in many materials 
from transition metals 
like NiGa$_2$S$_4$\cite{nakatsuji}, Na$_x$CoO$_2$\cite{naco} 
and YFe$_2$O$_4$\cite{nagano} 
to organic solids, $\kappa$-ET$_2$Cu$_2$(CN)$_3$\cite{kappa}. 
Theoretical studies on frustrated 
spin systems are conventionally intented on 
a search for "spin liquids"\cite{anderson} 
which generally lacks any long range order. 
However, although the magnetic order is destroyed by frustration 
another non-magnetic type of order often appears instead. 
Typical example is a dimer formation\cite{miyahara} 
and recently, a new gapless state called nematic order is found 
in the vicinity of ferromagnetic phase\cite{lauhili,nic}. 
\par
Charge degrees of freedom under frustration has also been 
an important topic from the past\cite{verway}, 
and one of the latest topics is a melting of charge order(CO) into 
the metallic state on the triangular lattice\cite{merino} 
in analogy with the "spin liquid". 
Although this seems to support an "orderless charge liquid" picture, 
the possibility of another type of ordering still remains unclear. 
In this Letter, 
we propose a new type of {\it partially ordered charge liquid} 
in such triangular lattice system. 
This liquid is nobel in that a charge degrees of freedom shows 
a coexistence of "statically ordered" and "liquid" components, 
where a gapless spatial long range order is possibly realized. 
The results are to be compared with 
an inhomogeneous metallic state found in 
$\theta$-ET$_2$CsZn(SCN)$_4$\cite{terasaki}. 
\par
\begin{figure}[t]
\begin{center}
\includegraphics[width=8cm]{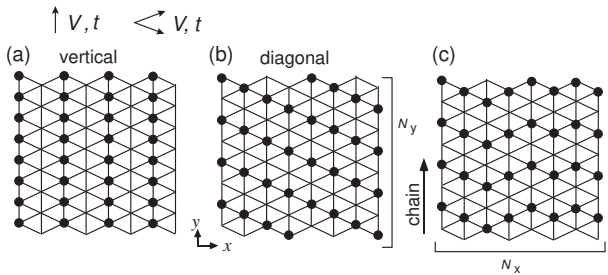}
\end{center}
\caption{
Schematic description of the typical CO on a anisotropic triangular 
lattice. 
The staggard alignment of charges in one particular direction(CO chain) 
with either (0101) or (1010)-type of configuration 
is present. 
Representative regular stacking of CO chains are 
(a) vertical and (b) diagonal striped states. 
One example of the random alignment is given in (c). 
}
\label{f1}
\end{figure}

One of the typical cases that CO resides is 
the quarter-filled electronic systems 
under strong electronic interactions\cite{chemrev}. 
It is conventionally described by the extended Hubbard model(EHM) 
including the on-site and inter-site Coulomb interactions, 
$U$ and $V$, respectively\cite{seo} besides the transfer integrals, $t$. 
Since the spin degrees of freedom plays only a secondary 
role when $t < V \ll U$, i.e. $J \sim t^2/U \ll t$, 
treating the spinless fermions at half-filling, $\rho$=0.5, 
is enough to understand the CO phenomena 
realized in the temperature range of $J < T < T_{\rm CO}$, where 
$T_{\rm CO}$ denotes the onset temperature of CO. 
We introduce a Hamiltonian of such $t$-$V$ model as, 
\begin{equation}
{\cal H}_{t-V} = \sum_{\langle i,j \rangle} \Big(
         -t_{ij}c^\dagger_i c_j + {\rm h.c.} + V_{ij} n_i n_j\Big). 
\label{tvham}
\end{equation}
Here, $c_j$ denotes the annihilation operator of fermions and 
$n_j(=c^\dagger_j c_j)$ is its number operator. 
The index $\langle ij \rangle$ are the nearest neighbor(nn) pair sites.  
We deal with the triangular lattice with anisotropy in one of three
directions. The geometry is reflected in $t,t'$ and $V,V'$ 
as shown in Fig.\ref{f1}, where the dashed values are 
the anisotropic ones. 
\par
\begin{figure}[t]
\begin{center}
\includegraphics[width=7cm]{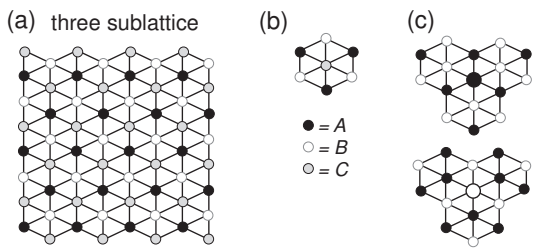}
\end{center}
\caption{
(a) Three sublattice structure; 
two sublattices($A$ and $B$) have the 0 and 1 configurations, respectively, while the 
third one($C$) could individually chose either 0 or 1. 
(b) Hexagonal unit of the three sublattice structure. 
The center site ($C$) is free, surrounded by the staggard edges. 
(c) Represetative contingents. 
When the three nearest $C$-sites 
happend to have the same configuration,  
the center site($A$ or $B$ in big circle) can switch its number without 
the energy loss. 
}
\label{f2}
\end{figure}

We first take the classical limit, $t=t'=0$, for simplicity, 
which exactly corresponds to the Ising $S$=1/2 spin system. 
According to Wannier, 
the ground state of the regular triangle, $V=V'$, 
is macroscopically degenerate and disordered, 
which are classified into two groups\cite{exact}. 
One group has the staggard alignment of 
charges (0101$\cdots$) in one direction which stack to 
form regular as well as random stripes shown in Fig.~\ref{f1}. 
Here, 0 and 1 denote the absence and the presence of charges, respectively. 
The other states are basically made up of three sublattices 
as shown in Fig.~\ref{f2}(a). 
Two out of three sublattices($A$ and $B$) have 
the fixed population(1 and 0) 
while each third sublattice site ($C$) can be arbitrary. 
Thus the system is piled up by the hexagon unit in Fig.~\ref{f2}(b), 
whose binding energy is always $-3V$. 
Besides the regular three sublattices, there are numbers of 
contingent freedom shown in Fig.~\ref{f2}(c). 
Since the energy of the three sublattice states 
including these contingents does not depend on the filling within 
$1/3 \leq \rho \leq 2/3$, the system is gapless at $\rho=0.5$\cite{expln}. 
This is in sharp contrast to 
to the striped states with the excitation gap of $2V$. 
\par
\begin{figure}[t]
\begin{center}
\includegraphics[width=7.5cm]{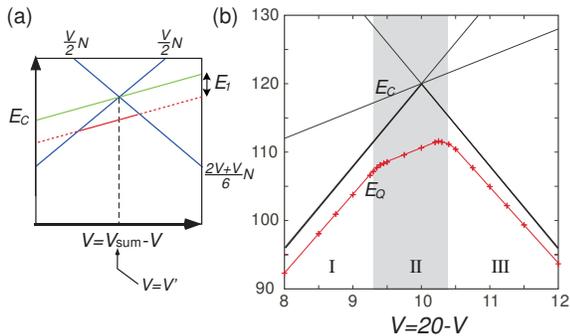}
\end{center}
\caption{
(a) Classical binding energy, $E_C$, 
at the fixed value of $V+V'=V_{\rm sum}$ as a function of $V$. 
The vertical and disordered striped states are realized at 
$V<V'$ and $V>V'$, respectively, while at $V=V'$, the three sublattice 
states with $E_C=(2V+V')N/6$ becomes degenerate with these stripes. 
Energy correction, $E_1$, up to first order perturbation of $t$ 
lowers the energy of the three sublattice state. 
(b) Ground state energy $E_Q$ of eq.(\ref{tvham}) at $N$=24 
with $V_{\rm sum}=20$ and $t=t'=1$. 
The diagram is classified into three regions; (I)vertical stripe, 
(II)pinball liquid, and (III)disordered stripe, 
where $E_Q$ behaves almost linear to respective $E_C$'s. 
}
\label{f3}
\end{figure}

When the anisotropy is introduced, 
the macroscopic degeneracy is lifted. 
At $V>V'$, the vertical stripe type in Fig.~\ref{f1}(a) becomes 
a unique ground state. 
As for $V<V'$, the charges align in a staggered manner 
along the anisotropic direction, e.g. as in Fig.~\ref{f1}(c). 
Since the each chain can choose either (0101) or (1010)-type of 
alignment, there remains a semi-macroscopic $2^{N_x}$-fold degeneracy, 
and the disorder survives in the $x$-direction. 
The classical binding energy, $E_C$, 
of the anisotropic triangle under the fixed value of $V+V'\equiv V_{\rm sum}$ 
is given in Fig.~\ref{f3}(a). 
The energy line follows 
$E_C \propto VN/2$ and  $-VN/2$ at $V<V'$ and $V>V'$, respectively, 
and equals exactly with the three sublattice ones at $V=V'$. 
\par
\par
In the highly frustrated disordered region an introduction of 
small $t_{ij}$ is expected to lift the degeneracy. 
Let us start from the regular triangle at $V=V' \gg t=t'$. 
The first term of eq.(\ref{tvham}) is considered as perturbations. 
In the striped states, the first order correction is absent 
and the energy changes to $\sim (V\!-\!t^2/V)N/2$. 
Since this effect is small, the stripes are quite robust. 
\par
\begin{figure}[t]
\begin{center}
\includegraphics[width=6cm]{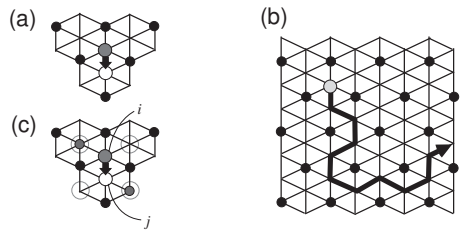}
\end{center}
\vspace{-5mm}
\caption{
(a) Representative basis of the three sublattice states near $\rho=1/3$
that mixes in the first order of $t$ at $\rho=1/3$ with one extra particle. 
(b) Schematic illustration of a pinball model. 
(c) Representative basis away from $\rho=1/3$, that is allowed to mix by $t$. 
The "ball" can move from $i$-th to $j$-th site when the extra two neighboring sites 
of $i$-th site with big circles together have the same number of balls with those 
of the $j$-th site. 
}
\label{f4}
\end{figure}

In contrast, the three sublattice states are modified at 
the first order level. 
For demonstration, we first deal with the simplest $\rho=1/3$ case. 
The charges align equally spaced like a Wigner crystal, 
e.g. in the $A$-sublattice. 
Next we add a single charge, and then 
there are $2N/3$ basis with the extra charge on one of the $B$ or $C$-sites. 
The charges can hop by $t$ to its neighboring $B$ or $C$-sites 
as in Fig.~\ref{f4}(a) without the loss of binding energy. 
Resultantly, 
the charge can move around avoiding the $A$-site just like a pinball, 
so we call this "a pinball liquid". 
The same situation holds at $\rho=2/3$ 
where we regard the "pins" as a hole-rich site after the particle-hole 
transformation. 
In the dilute limit, the extra charges thus form a honeycomb band. 
Such simple understanding is, however, non-tribial 
when the charge density becomes close to half-filling, $\rho \sim 0.5$, 
where about one-half of the $B$- and $C$-sites are filled with charges. 
The "balls" can at least move from one site to another following the rule shown in 
Fig.~\ref{f4}(c). 
Still, there are several issues to be examined. 
First problem is that a significant number of states including 
the contingents in Fig.~\ref{f2}(c) will prevail, 
which might mix by $t$ with regular three sublattice states 
and destroy the "pins". 
Another thing is that the free pinball picture no longer holds since there are 
interactions between balls themselves. 
Also, the validity of 
the above discussion is not clear 
when the anisotropy is introduced. 
\par
\begin{figure}[t]
\begin{center}
\includegraphics[width=7.5cm]{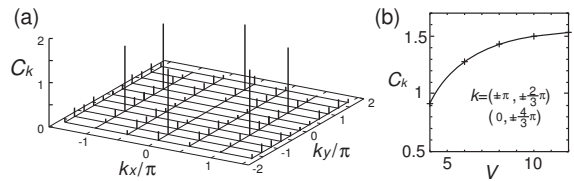}
\end{center}
\caption{
(a) Charge structural factor, $C_k$, at $V$=$V'$=10,$t$=$t'$=1. 
(b) $V$($=V'$)-dependence of the peak at $\vec{k}=(0,\pm
 4\pi/3),(\pm\pi,2\pi/3)$. 
}
\label{f5}
\end{figure}
\begin{figure}[t]
\begin{center}
\includegraphics[width=8cm]{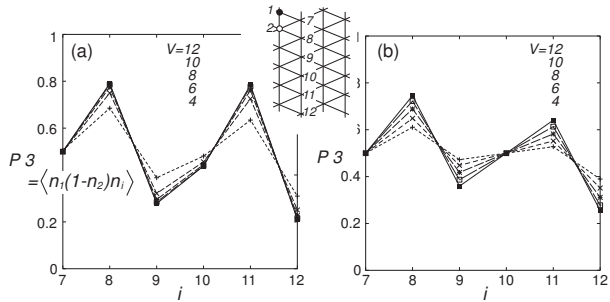}
\end{center}
\caption{
Three-body charge correlation function, $P_3(j)$, along the 
second chain($i=7-12$) at $t=t'=1$ for several choices of $V=V'$. 
Here, $i$ denotes the location of the site indicated in the inset. 
Those of the (a) the spinless fermion and (b) the hard core boson systems 
are shown. 
}
\label{f6}
\end{figure}

To clarify these points we proceed the numerical calculations on eq.(\ref{tvham}) at half-filling. 
We execute the exact diagonalization at $T$=0 on $N$=4$\times$6=24 cluster. 
Then {\it both the stripe type of 
two-fold states and the three sublattice states are compatible}. 
Figure~\ref{f5} shows the structural factor of the 
charge-charge correlation function at $V$=$V'$=10 and $t$=$t'$=1 
under the periodic boundary condition. 
It is defined as, 
$C_k=\frac{1}{N}\sum_{lm} \langle (n_l\!-\!1/2)(n_m\!-\!1/2)\rangle
 {\rm e}^{\rm i(l-m)k}$. 
The three sublattice type of 
peak structures at $\vec{k}=(\pm \pi, \pm 2\pi/3), (0,\pm 4\pi/3)$ are 
observed, which grow with increasing $V$. 
To see the details, we calculate 
the three-body correlation function, $P_3(j)\equiv \langle n_1 (1-n_2) n_j \rangle$, 
which reveals the population of the $j$-th site 
when the {\it 1}-st and the {\it 2}-nd site is present and absent, 
respectively. 
The ones along the $y$-direction for several choices of $V=V'$ are shown in 
Fig.~\ref{f6}(a). 
A clear three fold structure of $A$-$B$-$C$ type 
with different density, namely $P_3(A)+P_3(C) \sim 2 P_3(B) $, is present, 
suggesting the existence of particle-hole symmetry. 
Here, $P_3(A)$ denotes the correlation when $j$-th site belongs to the $A$-sublattice. 
Thus, although the amplitude of the "pins" are squeezed to $\sim 0.8$ 
due to the quantum effect, a characteristic "pinball"-type of 
three body correlation is present. 
It originates from the geometrical frustration 
and essentially differs from the two-body ones in the CO states. 
The correlation decays with distance so that we cannot 
determine whether it lasts in the bulk limit or not within our scheme. 
Nevertheless, the amplitude of ``pins'' are growing with $V$ to form 
a firm structure at least over considerable distances. 
\par
The pinballs sustain even when the anisotropy of 
$V\neq V'$ or $t\neq t'$ is introduced. 
Actually, in the case of $\rho=1/3$ and 2/3 with a single ``ball'', 
the only modification is the binding energy per "ball" from 
the previous $-3V$ to $-2V-V'$. 
If only the first order correction of energy by $t$, 
$E_1$, is finite the three sublattice state should overcome the others 
as in Fig.~\ref{f3}(a) and form a new phase at $V \sim V'$. 
Figure~\ref{f3}(b) shows the ground state energy of eq.(\ref{tvham}) 
together with $E_C$ at fixed $V\!+\!V'\!=\!20$. 
The diagram is separated into three regions; 
the energy line behaves parallel to different classical states, 
(I) vertical stripe, (II) three sublattice, and (III) disordered stripe. 
The pinball liquid realized in (II) extends at $V \ne V'$
over a width of order $t$. 
Actually, the first order energy gain $E_C-E_Q$ at $V=V'$ is several 
times larger than the second order gain in the (I) and (III) regions. 
In this way, the pinball liquid is quite dominant at around $V \sim V'$. 
\par
\begin{figure}[t]
\begin{center}
\includegraphics[width=8cm]{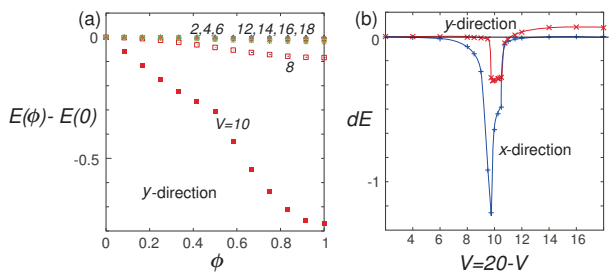}
\end{center}
\caption{
(a)Variation of the ground state energy, $E(\phi)-E(0)$, 
as a function of $\phi=0-\pi$, 
under the twisted boundary condition in the $y$-direction for 
several choices of $V$ with $V+V'=10$ fixed. 
(b) The amplitude of variation $E(\pi)-E(0)$ as a function of $V$ along 
$V+V'=10$, when the boundary is twisted in the $x$- and $y$-directions. 
Regions (I)-(III) correspond to those of Fig.\protect\ref{f3}
}
\label{f7}
\end{figure}

It is natural to expect that the gapless situation found in the 
classical three-sublattice states still holds when $t$ is introduced. 
Since the direct gap estimation under the finite system size 
is difficult and unreliable, 
we instead focus on the coherence length of each characteristic state. 
We twist the boundary condition as e$^{{\rm i} \phi}$, 
where $\phi=0$ and $\pi$ correspond to the periodic and anti-periodic boundary 
conditions, respectively. 
The ground state energy, $E(\phi)$, 
should have considerable $\phi$-dependence only when the coherence lasts 
longer than the system length. 
Figure~\ref{f7}(a) shows the variation of $E(\phi)-E(0)$ as 
a function of $\phi$ when twisting the boundary in the $y$-direction 
with $V+V'$ fixed. 
In the regions (I) and (III), 
$E(\phi)$ remains almost constant, 
reflecting the insulating character. 
In contrast, the ones in (II) show large $\phi$ dependence, 
indicating that pinball wave function has a delocalized character at least 
over dozens of sites. 
Figure~\ref{f7}(b) shows the amplitude of the energy variation, $E(\pi)-E(0)$, 
when twisting $\phi$ in the $x$ and $y$-directions. 
One actually finds a significant difference that separates (II) from other 
two states. 
\par
\par
In summary, we proposed the existence of a pinball liquid state on a triangular lattice  
in the spinless fermion systems 
with nn Coulomb interactions over a wide range of filling, 
$1/3 < \rho < 2/3$. 
It is an ordered liquid formed out of disorder 
by the lifting of macrscopic degeneracy 
and has a nobel three-body correlation. 
Thus, although the state is metallic 
it is driven by the local correlation of electrons at $V \gg t$ and 
is not related to the details of the Fermi surface, namely the 
geometrical structure of $t$. 
It is also insensitive to the anisotropy of $V \ne V'$ and survives 
quite away from the regular triangle. 
Further, it is expected to sustain when the degrees of freedom other than charges 
such as spins and orbitals are introduced. 
\par
This nobel liquid is quite universal. 
For example in the hard core boson systems (XXZ spin systems), 
$P_3(j)$ also shows a clear 
three-fold pinball character as given in Fig.~\ref{f6}(b). 
Therefore we conclude that it 
is not much influenced by the statistics and is also applicable to the 
large-$U$ region of EHM at quarter-filling. 
\par
Recently there had been other theoretical 
studies on the anisotropic triangular lattice that argues 
the existence of the three-fold states\cite{hwata,kaneko}. 
Unfortunately, their treatment based on the mean-field wave functions 
rather take account of the instability of the Fermi liquid 
towards the two-body correlations, 
and could not capture the essential features of the pinball liquid, 
namely the frustration induced three-body correlations. 
The present theory gives the explicit and clear-cut answers to 
the nobel nature of this system 
alluded to preliminarily in such previous studies\cite{merino,hwata,kaneko}. 
\par
Related experimental studies are explored 
on the CO materials, $\theta$-ET$_2X$. 
Recently, a coexistence of the short range diffuse spot of 
two-fold and three-fold periodicity is observed 
in the metallic states of materials containing the 
Rb\cite{wat1} and Cs\cite{wat2} ions, 
which have $V/V' \sim$ 0.85-0.9\cite{tmori}. 
From our strong coupling viewpoint, 
both two-fold striped and pinball liquid states are compatible 
(though not coexisting) 
without having special assumptions such as long range interactions 
or the particular shape of the Fermi surface. 
If the systems were in the critical region of 
these two competing characteristic states, 
some additional effects such as impurity, temperature(entropy effect), 
or electron-phonon interaction might induce 
a coexistence or domain structure 
that cope with these anomalous experimental findings\cite{terasaki}. 
\par
In conclusion, 
we found a new type of liquid state in the charge ordered system under 
geometrical frustration, which is possibly relevant at temperature range 
of $J < T \ll V$. 
In the strong coupling limit, 
the frustration among electronic interactions, $V$, 
drives the system to disorder. 
The introduction of small but finite $t$($\gg V$) transforms 
the system to a strongly correlated quantum liquid with 
three sublattice correlation, $A$-$B$-$C$, 
whose electronic density is, $n_A+n_B \sim 2n_C$. 
About half of the charges become a Wigner-crystal-solid 
while the rest remains a liquid. 
It is nobel in that 
an originally unique charge degrees of freedom 
spontaneously separates into the solid and liquid-like part 
and coexist in the same system. 
Such correlation induced order out of disorder 
distinctively differs from the usual charge ordering 
or charge-density-waves formed 
from the instability of the Fermi liquid. 
This state remains valid regardless of electronic filling, 
the shape of the Fermi surface, 
anisotropy of $V$, and also the statistics of the particles, 
and endows a new detailed insight on the topic of frustration. 
\par
We thank K. Kubo, S. Miyahara, I. Terasaki, R. Kondo and M. Watanabe for discussions. 


\begin{references}
\bibitem{nakatsuji}S. Nakatsuji, Y. Nambu, H. Tonomura, O. Sakai, S. Jonas, C.
Broholm, H. Tsunetsugu, Y. Qiu, and Y. Maeno, Science {\bf 309} (2005) 1697. 
\bibitem{naco} M. L. Foo, Y. Wang, S. Watauchi, H. W. Zandbergen, T. He, R. J. Cava, 
and N. P. Ong, Phys. Rev. Lett. {\bf 92} (2004) 247001. 
\bibitem{nagano} S. Funahashi, J. Akimitsu, K. Siratori, N. Kimizuka, M. Tanaka, 
H. Fujishita, J. Phys. Soc. Jpn. {\bf 53} (1984) 2688. 
\bibitem{kappa}Y. Shimizu, K. Miyagawa, K. Kanoda, M Maesato and G. Saito, Phys. Rev. Lett.
{\bf 91} (2003)107001. 
\bibitem{anderson} P. W. Anderson, Mater. Res. Bull. {\bf 8} (1973) 153; 
P. Fazekas and P. W. Anderson, Philos. Mag. {\bf 30} (1974) 423.
\bibitem{miyahara}S. Miyahara, K. Ueda, Phys. Rev. Lett. {\bf 82} (1991) 3701. 
\bibitem{lauhili}A. Laeuchli, J. C. Domenge, C. Lhuillier, P. Sindzingre, M. Troyer, 
Phys. Rev. Lett. {\bf 95} (2005) 137206.
\bibitem{nic}N. Shannon, T. Momoi, P. Sindzingre, Phys. Rev. Lett. {\bf 96} (2006) 027213.
\bibitem{verway}E. J. W. Verway, Nature {\bf 144} 327 (1939). 
\bibitem{merino}J. Merino, H. Seo, M. Ogata, Phys. Rev. B {\bf 71} (2005) 125111.
\bibitem{terasaki}F. Sawano, I. Terasaki, H. Mori, T. Mori, M. Watanabe, 
N. Ikeda, Y. Nogami, Y. Noda, Nature {\bf 437} (2005) 522.
\bibitem{chemrev} H. Seo, C. Hotta, H. Fukuyama, Chem. Rev. {\bf 104} (2004) 8558, 
and the references therein. 
\bibitem{seo} H. Seo and H. Fukuyama, J. Phys. Soc. Jpn. {\bf 66} (1997) 1249.
\bibitem{exact} G. H. Wannier, Phys. Rev. {\bf 79} (1950) 357.
\bibitem{expln} To be exact, the grand canonical Hamiltonian, ${\cal H}-\mu N_e$, 
with $\mu=3V$ and tne electrnon number, $N_e$, denotes the gapless picture 
of the three sublattice states at $1/3 \leq N_e/N \leq 2/3$ 
in the classical limit. 
\bibitem{wat1}M. Watanabe,Y. Nogami, K. Oshima, H. Mori, S. Tanaka, 
J. Phys. Soc. Jpn. {\bf 68} (1999) 2654. 
\bibitem{wat2}M. Watanabe, Y. Noda, Y. Nogami, H. Mori, J. Phys. Soc. Jpn. {\bf 73} (2004) 116; 
M. Watanabe, Y. Noda, Y. Nogami, H. Mori, J. Phys. Soc. Jpn. {\bf 74} (2005) 2011.
\bibitem{hwata} H. Watanabe and M. Ogata, condmat/0602536/.
\bibitem{kaneko}  Kaneko and M. Ogata, condmat/0602536/.
\bibitem{tmori} T. Mori, Bull. Chem. Soc. Jpn. {\bf 73} (2000) 2243. 
\end{references}
\end{document}